# Optimal Indoor AP Placement: A Case Study

Farah Natiq Qassabbashi, Qutaiba Ibrahem Ali, *Member, IEEE*

*Abstract*— **Wireless networks in a room are strongly affected by interferences. To alleviate these effects and enhance the performance of the wireless networks, some optimization was carried out. In this work, an analytical study was introduced to determine the optimal number of access points with their positions on the ground floor at the Architecture Engineering department building - University of Mosul. The implementation has been done using a web-based Wi-Fi and IoT design tool called Hamina Network Planner, and a Wi-Fi Network Planning and Site Survey Software called NetSpot. The experimental results show that the simulation values of the available APs are approximately matched with the real time manual values, achieving the best rates of -22dBm and -31dBm respectively. However, the number of currently available Access Points is not sufficient to cover the building area, so that two scenarios were suggested to overcome this issue. In the first scenario, two access points have been added at different positions in the building depending on the Hamina Network Planner, and in the second one, the transmitted power has been increased. The simulation results demonstrate that the overall coverage rate enhanced and can include most of the building area.**

*Index Terms*— **AP Placement, Network Design, Optimization methods, Coverage, Indoor Map, Wi-Fi.**

## I. INTRODUCTION

THE rapid development of wireless communication systems has simplified to obtain information in many real-time applications. However, offering full coverage for each user within the design space is the primary aim of Wireless Local Area Network (WLAN) deployment. The number of Access Points APs positioned inside the design area affects coverage. A high number of APs increases deployment costs, and few of them leads to coverage gaps that prevent client's access to their data via APs [1].

In order to determine the optimum numbers & placing of APs, some studies have used radio frequency (RF) site surveys [2]. Others use mathematical modeling and optimization methods [1] [2], where the region is divided onto rectangles and the APs are positioned in the centers of every rectangle. The grid size must to be small sufficient so that the problem's dimension is very high. Other authors prefer continuous equations where the positioning of APs is free.

The indoor space may be divided into various parts, using nodes spread over the floor. As soon as fashion modeling the network, all nodes are assumed to be unmoved, and we have to be taken into consideration the access points loss of power as a function of the distance, obstruction, as well as interference in signals from energy reflection . The kind of material used to construct the obstacle impacts what amount of signal attenuation is caused by it [3]. Whenever the node is throughout a barrier constructed with aluminum or glass, the coefficient of absorption will typically be greater than when it is within a wall made of bricks and cement. The absorption coefficient additionally varies based on how thick a wall is. Since the walls and ceilings reflect signals, the signal strength at every node depends on the reflection coefficient and the phase that both direct and reflected signals arrive at the receiver. Therefore, some researches are investigating ideas into how to organize Wi-Fi access points to ensure that everybody in a city receives up a strong signal via at least three reference APs while maintaining the number of required APs as few as possible [4].

There are two different approaches to perform wireless site surveys: manual and virtual. In the first method, multiple assigned access points are utilized for verifying access point places, then a site survey program monitors wireless signals to locate access points and estimate signal strength. Using simulation tools, users can construct virtual access points in a framework for a visual survey, predict expected coverage, and modify the total number and position of access points [5]. Such tools could provide clients highly realistic solutions through taking into account realistic radio propagation characteristics for planning the site, like selecting an antenna, band (2.4 GHz vs. 5 GHz), and evaluation of the environment (obstructions, wall materials, or building ceiling height).

To describe the optimization problem, some investigators employed path loss, whereas others made use of power at the receiver [2]. For identifying the best placement for APs, we follow the second approach during this study. Power at the receiving end will be used to calculate the AP's coverage area due to the loss of signal strength forced on by obstacles and distance across the transmitter and receiver.

In this paper, advanced graphic technology is leveraged to allow explicit representation of complex map details for the Architecture Engineering department building - University of Mosul that can be utilized for optimal AP placement for more accurate location usin Hamina Network Planner tool and NetSpot software planner.

The paper is organize as follows: Section II shows the literature review. Section III presents a case study at the Architecture Engineering department building - University of Mosul. The results and discussion is described in Section IV. Finally, section V. summarizes the paper and suggests some of the future works.

## II. LITERATURE REVIEW

This section includes a detailed review of the approaches and methodologies for problems related to the positioning and density of APs in WLANs. The issue is to acquire maximum coverage, minimal interference and improved network performance as shown in TABLE I.

TABLE I
SUMMARY OF Optimization ACCESS POINT PLACEMENT

| Paper | Problem | Research methodology | Suggested solutions |
|---|---|---|---|
| S. Kouhbor et al. [1]2005 | Find the optimum Aps number and placement onto the WLAN. | 1- A model constructed using mathematical concepts. 2-Discrete Gradient optimization algorithm. 3-Cisco AP models and IEEE 802.11b standards for testing. 4-Path loss model for develop the optimization problem | 1-The size of design area and user distribution affect AP placement and coverage. 2- Path loss increases as distance increase, confirming the expected behavior of attenuation. |
| S. Kouhbor et al. [2]2006 | Find the optimum number of APs and positions within an obstacle-filled location. | 1-A novel mathematical model. 2-A global optimization (AGOP) algorithm. 3- Cisco Aironet 1100 Series for testing the coverage model. 4-Path loss model for develop the optimization problem | 1- This model has the capability of overcoming WLAN coverage issues for a design region based on different user counts, buildings sizes, user places, and power levels that affect AP requirement. 2- More APs are needed as power decreases |
| S. F. Rodd et al. [3]2009 | Compute the ideal number and placement of APs while considering reflections, obstructions, and signals deterioration. | 1. The method of discrete gradients. 2. The genetic method. 3. The technique of global optimization. 4- Heuristic Search Technique. | Heuristic Search Technique is the best one and can provide faster and accurate solution, by moving the AP diagonally depending on signal strength estimation. |
| K. Farkas et al. [4]2013 | Obtain the optimum number and location of APs in indoor positioning. | 1- A simulated annealing based algorithm. 2- New simulation tool in MATLAB environment. 3- ITU indoor wireless signal propagation model. | 1- Receive the signal from a minimum of three reference APs in the selected indoor area. 2- The method has O(n) complexity. 3- Keep the number of Aps as low as possible. 4- The overall cost decrease |
| G. Lee [5]2015 | Addres the limitations of traditional two-dimensional facility location models | 1-A novel 3D coverage location model. 2-Model parameters includes Gaussian random variable, path loss exponent, and initial signal strength. For applications in the real world, a coverage model additionally includes radio signal propagation. | 1- Incorporating the vertical dimension in multi-story buildings. 2- More efficient calculations in determining AP placements. 3- More manageable and practical for real-world applications. |
| X. Du [6]2017 | An interior map system that assists in AP placement by providing a graphical representation and coordinate system | 1- Vector graphic indoor map system. 2- A wall detection algorithm. 3- Particle swarm optimization (PSO). 4- k-nearest neighbors positioning algorithm. | 1. Higher positioning accurate can be accomplished through map-assisted AP placement. 2. It significantly decreases the maximum error distance. |
| Y. Tian [7]2018 | AP positioning optimization challenge for both the coverage and the localization | 1- Lower bound of Cramer-Rao CRLB that Provide the AP placement techniques. 2- A heuristic genetic algorithm. 3- Motley-Keenan model. | 1. In comparison to the other three commonly employed approaches, this one minimizes time. 2. By employing the simulated annealing procedure and fingerprinting difference, it achieves nearly ideal performance. |
| M. P. Fawzan [8]2018 | An excellent access point in the Law and Shari'ah Faculty Building | 1- The method of Bayesian probability. 2- Using manual random sampling for estimating the signal's degree. | The proposed method is successful and can reduce the blank spot area. |
| M. M. Abdulwahid [9]2019 | optimal AP placement for Indoor environment | 1. A two-step algorithm for location. 2. The Received Signal Strength (RSS) was calculated via NetSpot Pro software. 3- A MATLAB program with a GUI was used for the localization phase | When selecting the most suitable position, the effects of overlapping and co-channel interference have been taken into account. |
| P. Imputato [10]2024 | Determining the number of AP groupings that may sent simultaneously and successfully | 1- A coordinated Spatial Reuse c-SR interference model. 2- They use ns-3 simulations to evaluate the performance. | 1. Increasing a dense system's throughput due to 2.3 times 2. Significantly decreased the Head of Line delay. 3. This study could be the first to employ realistic ns-3 simulation to assess a c-SR proposal's performance. |
| R. Arunkumar [11]2024 | To optimize network thyroughput, optimized access point selecting (APS) and the standard load balancing (LB) technique development. | 1. The combination of wireless (WiFi) and light (Lifi) fidelity. 2. (Deep LSTM) Deep Long Short Term Memory was employed to perform location prediction 3- Sewing Training Inspired Optimization (STIO). 4. The data is going to be traveled via the Multipath Transmission Control Protocol (MPTC). | 1. A smaller latency 0.113 ms . 2-The high energy efficiency of 0.097 Mbits/joules. 3. A minor handover occurrence of 6.100. 3. An excellent network throughput 0.905 . |

.

## III. CASE STUDY: ARCHITECTURE ENGINEERING BUILDING AT UNIVERSITY OF MOSUL: A SITE SURVEY

### A. Research location and purposes

The Architecture departments building at the College of Engineering is located within the University of Mosul, at 43.135054 Longitude and 36.387076 Latitude, with approximately building area of 3273 m2 as shown in Fig.1. This campus, since its formation, it adopts wireless networks whereby the academic community gains easy access to the internet at any given area, approximately 75 members of the college use the network on a daily basis. However still a below par network because there are certain areas that are not covered by network (blank spot). For this reason, a study must be done to determine the position and number of AP that is capable of covering this building effectively with Wi-Fi network by making use of computer-based application software. For more simplicity, the indoor location in the ground floor was divided into 19 areas, as shown in Fig.2.

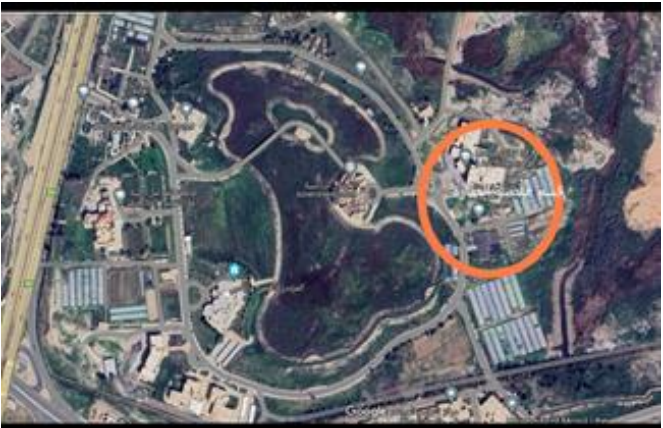

Fig. 1. Architecture Engineering Department building location inside the University of Mosul.

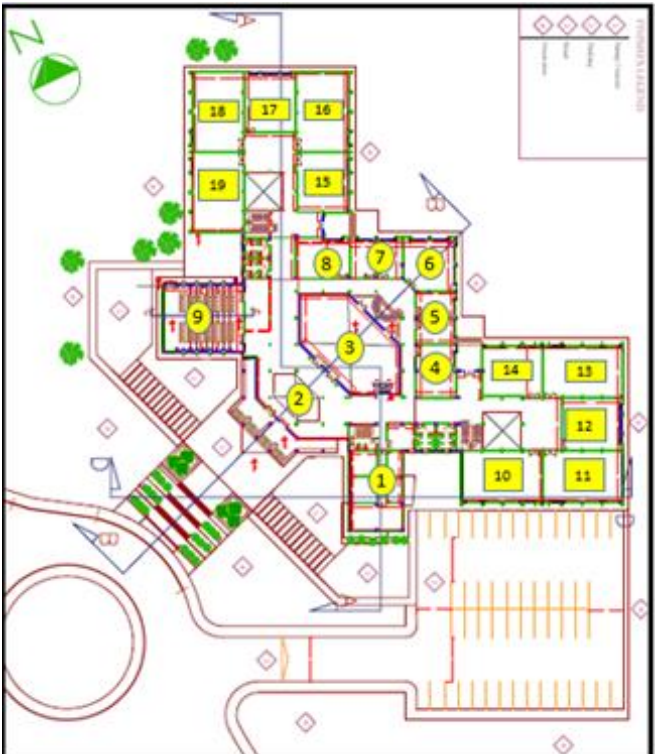

Fig. 2. Architecture Engineering Department building Map (Ground Floor).

### B. Research method

The research method used in this study is to analyze the network architecture in the Department of Architecture Engineering building regarding Access Points. To determine the placement of Access Points with high probability values at the most important parts of the ground floor, a web-based Wi-Fi and IoT design tool called Hamina network planner was used [12], so that the APs are located in the same position in the building area of Architecture Engineering. This tool is totally cloud-based and the free trial allow to use only three Access points to predict the best position in the installation at the given area; so that the Wi-Fi signal is optimally received by all the users [13]. To obtain the primary data, data retrieval is done using manual random sampling according to the chosen areas using Netspot software to measure the power of the Signal in each area as shown in Fig.2. So that a comparison between predicted values and real values can be done [14].

To start with Hamina network planner; a floor plan must be uploaded and then an accurate scale must be chosen, since this planner uses distance to predict how far signal will travel from each access point. Attenuating Objects such as walls and doors must be drawn, because they create obstacles that reduce the signal strength. To use the Hamina's heatmaps the Technology and Band for WiFi must be chosen depending on requirements [15]. To add Access Points, the technology, vendor, model, mounting type and height must be selected, as well as additional options such as Access Point Name, Transmit Power settings, connected via Ethernet selector, and Power allocation from switch must be selected, as shown in Fig.3.

Technology:

Wi-Fi

Access point:

Ubiquiti        AC Long-Range

Mounting type:        Height:

Ceiling        2.7    m

Access Point Name:

Access Point 1

Channels:

Auto-channel assignment is **enabled**.
Channel settings

Transmit power:

2.4 GHz        5 GHz

24   dBm        22   dBm

11 / 20 MHz        52 / 80 MHz

Fig.3. Hamina heatmaps options

### C. Research implementation

Research implementation consist of Two scenarios, the first one is to generate Hamina's heatmaps and decide the suitable position of the Access points that Corresponds to the same location in the Architecture Engineering building, and specify the prediction power in each area[16], as shown in Fig.4.

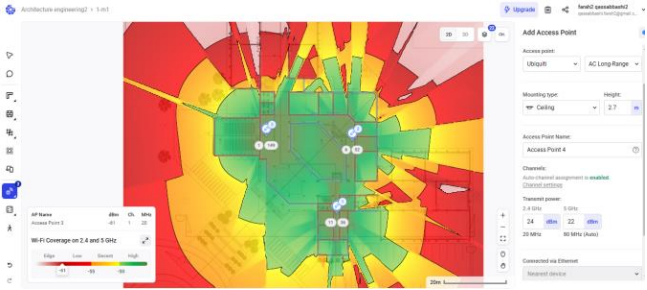

Fig.4. Hamina's heatmaps for the first scenario

The second scenario can be done through a real time measurements using Netspot software to obtain the power of the signal in each area [17], as shown in Fig5.

A comparison between the predicted values of Hamina network planner and the real time values of Netspot inspector can be shown in Fig.6.

In the third scenario Fig.7, the power of adding two APs is estimated, simulated results and Heatmaps indicate that the coverage of the intended areas increases [18]. Fig.8 and Fig.9 demonstrates the Heatmaps of the first and third scenarios in more details.

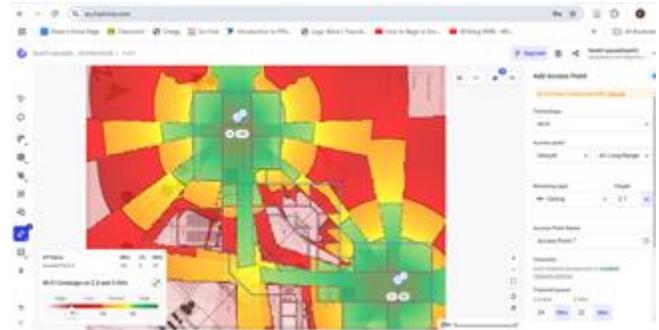

Fig.7. the third scenario heatmap.

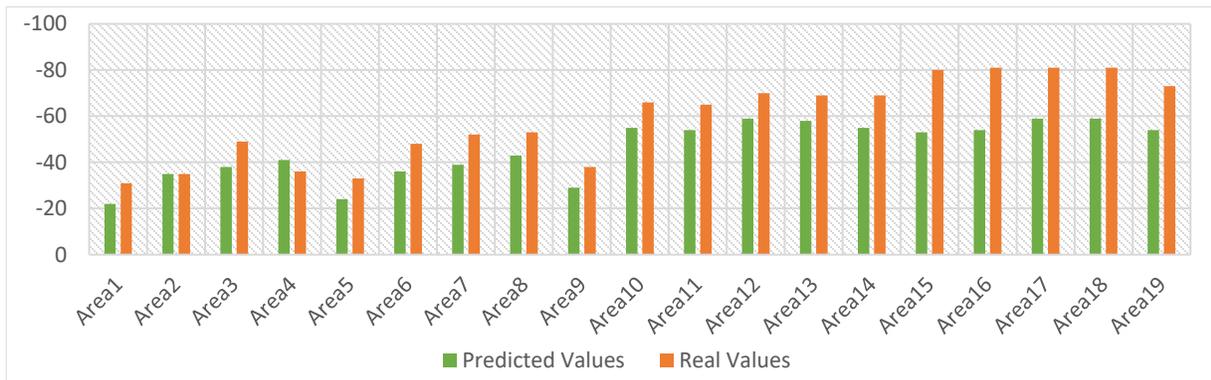

Fig. 5. Netspot software measurement.

Fig. 6. Best power comparison between Hamina network planner and Netspot inspector

## IV. RESULTS AND DISCUSSION

The outcomes obtained via the literature review as well as system analysis involves the achievement of the optimum access point location for the user to get the Wi-Fi signal as effectively as possible. Simulation and real time results indicate that the maximum power was obtained at 1,2,4,5 and 9 areas, where the access points are nearly located. In contrast to 10, 11, 12, 13, 14, 15, 16, 17, 18 and 19, that demonstrates the worst power value, so that the coverage is too low. For this purpose, another two scenarios have been taken into account to obtain an optimize coverage for these areas.

In the fourth scenario, the transmitted power have been increased, so that the overall building coverage is achieved as shown in Fig.10.

In this case, the estimated transmitted power using Hamina Network Planner tool of the first AP at 2.4 GHz and 5GHz bands will be 26dBm and 22dBm respectively. In addition, for the second AP will be 38dBm and 22dBm, and finally for the third AP will be 45dBm and 22dBm respectively.

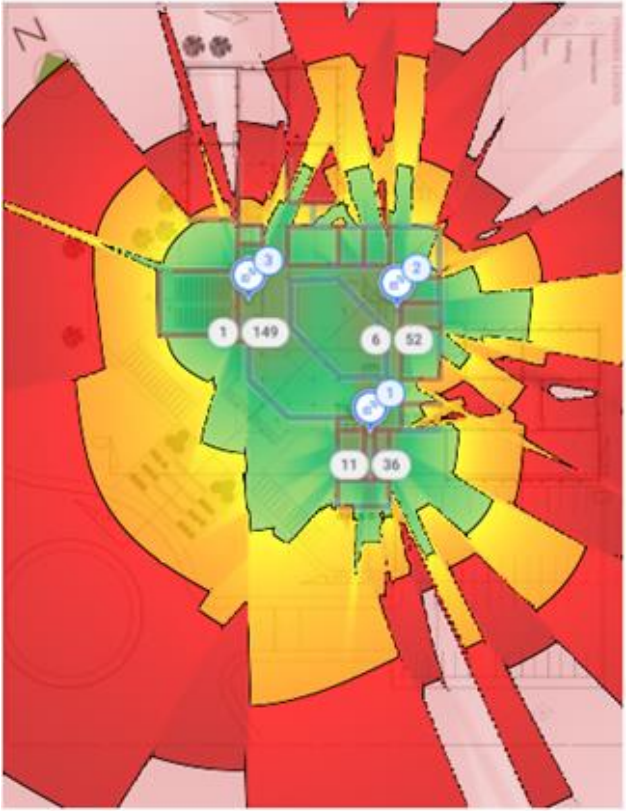

Fig.8. The first scenario heatmap

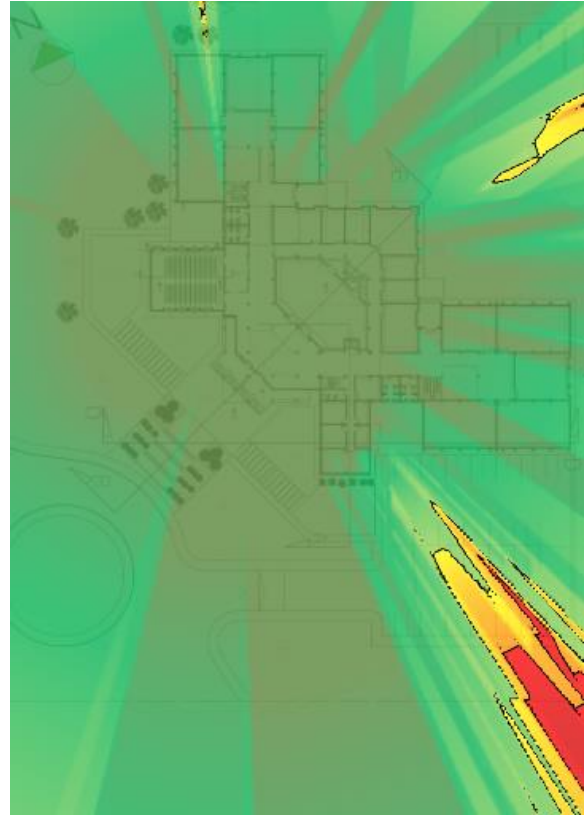

Fig.10. The fourth scenario heatmap

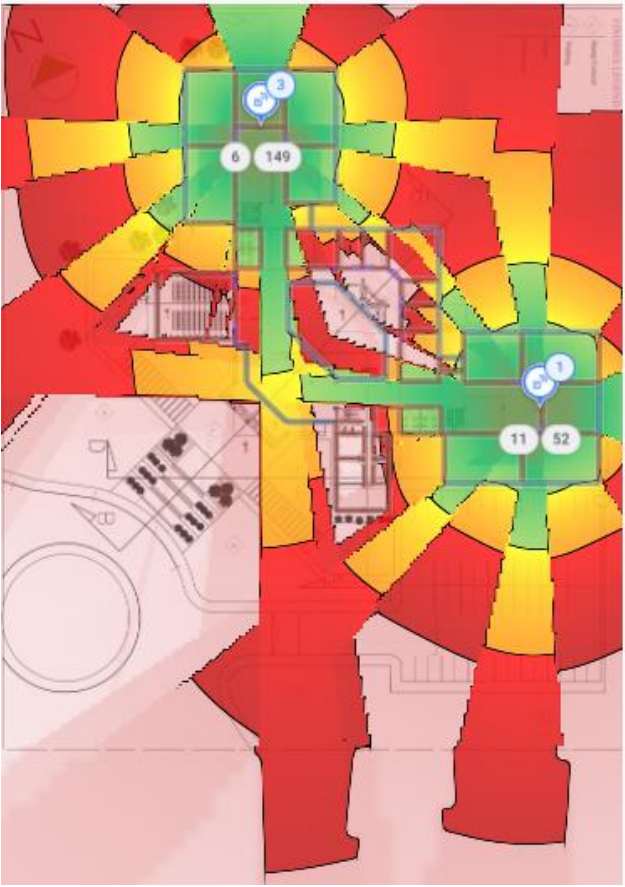

Fig.9. The third scenario heatmap

## V. CONCLUSION AND FUTURE WORKS

Effective planning is necessary to any successful wireless AP deployment, since it involves a set of goals and requirements that must be met. The goal of this work is to determine the optimal number and best position for AP placement based on realistic measuring, that can improve the overall coverage of the Architecture Engineering department building - University of Mosul. In this analytical study, Hamina Network Planner design tool was used to estimate the signal strength and power values at different areas in the building. Furthermore, NetSpot Software used to evaluate the real time power in the intended areas. The experimental results indicate that the simulated values approximately matched with the real values, achieving best rates of -22dBm and -31dBm respectively. However, the used access points are still insufficient to cover the building's total area. In order to increase the rate of coverage for the majority of the building's areas, two additional access points have been proposed in the third scenario, and the transmitted power have been increased in the fourth scenario so that the overall coverage is approximately achieved. For the future work, an AI Robotic Access Point can be used, so that it can provide wireless connectivity through autonomously move around a building.

## VI. ACKNOWLEDGEMENT

The authors would like to thank University of Mosul (https://www.uomosul.edu.iq/), Mosul, Iraq and Architecture

Engineering Department-College of Engineering for their support in the present work.

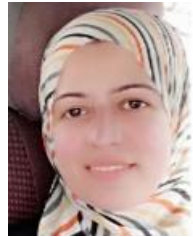

**Farah Natiq Qassabbashi** graduated with a (B.Sc.) degree in 2004 from College of Engineering, Department of Computer Engineering, University of Mosul. She received his M.Sc. degree in Computer Engineering in 2021 from College of Engineering, Computer Engineering Department, University of Mosul and followed this nomination as an assistant lecturer at the same university. She has published four papers in international scientific journals and international scientific conferences. Her research areas of interest include computer network, design and analysis of dependable hardware architectures, fault-tolerance, VLSI, and FPGA-based digitalsystems. She can be contacted at email: farah.qassabbashi@uomosul.edu.iq.

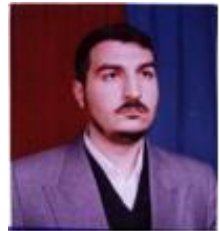

**Qutaiba Ibrahem Ali** received the B.S. and M.S. degrees from the Department of Electrical Engineering, University of Mosul, Iraq, in 1996 and 1999, respectively. He received his Ph.D. degree (with honour) from the Computer Engineering Department, University of Mosul, Iraq, in 2006. Since 2000, he has been with the Department of Computer Engineering, Mosul University, Mosul, Iraq, where he is currently a lecturer. His research interests include computer networks analysis and design, real time networks and systems, embedded network devices and network security and managements. Dr. Ali has attended and participates in many scientific activities and gets awards for his active contribution and has 161 published papers.